\documentclass[ijoc,nonblindrev]{informs3} 

\OneAndAHalfSpacedXI

\usepackage{natbib}
 \bibpunct[, ]{(}{)}{,}{a}{}{,}%
 \def\bibfont{\small}%
 \def\newblock{\ }%
\TheoremsNumberedThrough     

\EquationsNumberedThrough    

\usepackage{graphicx,amssymb,amsmath,textcomp}
\usepackage{color,textcomp}
\usepackage[bookmarks=false]{hyperref}
\hypersetup{pdfstartview={FitH}}
\usepackage[letterpaper]{geometry}
\usepackage[official]{eurosym}
\usepackage{bm}

\usepackage{url}
\usepackage{float}
\floatstyle{ruled}
\newfloat{model}{thp}{lop}
\floatname{model}{Model}

\usepackage{algorithm2e,cases}

\def\rit{\mathbb{R}}

\begin{document}
\RUNAUTHOR{Borraz-S\'anchez et al.} 
 \RUNTITLE{Convex Relaxations for Gas Expansion Planning}
 
\TITLE{Convex Relaxations for Gas Expansion Planning}
\ARTICLEAUTHORS{
\AUTHOR{Conrado~Borraz-S\'anchez, Russell~Bent, Scott~Backhaus}
\AFF{DSA-4: Energy \& Infrastructure Analysis, LANL, Los Alamos, NM-87545, USA}


\AUTHOR{Hassan~Hijazi, Pascal~Van~Hentenryck}

\AFF{NICTA and ANU, Canberra, 260, Australia.}
}
\ABSTRACT{

  Expansion of natural gas networks is a critical process involving
  substantial capital expenditures with complex decision-support
  requirements. Given the non-convex nature of gas transmission
  constraints, global optimality and infeasibility guarantees can only
  be offered by global optimisation approaches. Unfortunately,
  state-of-the-art global optimisation solvers are unable to scale up
  to real-world size instances.  In this study, we present a convex
  mixed-integer second-order cone relaxation for the gas expansion
  planning problem under steady-state conditions. The underlying model
  offers tight lower bounds with high computational efficiency. In
  addition, the optimal solution of the relaxation can often be used
  to derive high-quality solutions to the original problem, leading to
  provably tight optimality gaps and, in some cases, global optimal
  solutions. The convex relaxation is based on a few key ideas,
  including the introduction of flux direction variables, exact
  McCormick relaxations, on/off constraints, and integer
  cuts. Numerical experiments are conducted on the traditional Belgian
  gas network, as well as other real larger networks. The results
  demonstrate both the accuracy and computational speed of the
  relaxation and its ability to produce high-quality solutions.  
}


\maketitle	

\section{Introduction} 
\label{sec:Introduction}

In recent years, the construction of natural gas pipelines has
witnessed a tremendous growth on a world-wide level. In the U.S., for
instance, a \$3 billion expansion project of the gas pipeline system
in New England is planned for late 2016. In Europe, the European
Investment Bank is supporting a \euro{}98 million project for the
expansion of gas pipelines in western Poland, to be completed by 2017.
These expansion projects aim at increasing gas flow capacity on
existing pipeline systems and/or bringing new gas wells into
production and commercialization. In addition, the expansion or
reinforcement of a pipeline network can also be considered as a
risk-awareness strategy to fulfill short or long-term operational
management requirements when unforeseen events occur such as component
failures or excessive stress and congestion due to extreme weather
conditions. These events were observed in New England during the polar
vortex experienced in January 2014, when major gas-fired power plants
in the northeast of the U.S. were forced to shut down due to
mechanical problems and shortages of gas fuel supplies, which drove
wholesale power prices up 

According to the U.S. Energy Information Administration (EIA), a
project for the development and expansion of a Gas Transmission
Network (GTN) takes an average of three years from its first
announcement until its completion~\citep{Ref_EIA2008}. The project
starts by determining the market needs within an open season exercise
where nonbinding agreements of capacity rights are offered to
potential customers. The second step consists in developing the
expansion design with initial financial commitments from the potential
customers. Note that expansions of the gas system include the
installation parallel pipelines along existing ones (looping), the
conversion of oil pipelines to natural gas pipelines, or the
reinforcement of specific pipeline sections.

In this paper, we address the Gas Transmission Network Expansion
Planning (GTNEP) problem where the goal is to fulfill projected future
gas contracts and to increase the reliability of a gas system under
steady-state conditions. A Mixed-Integer NonLinear Programming (MINLP)
formulation is proposed to model the design requirements and minimize
expansion costs. Given the non-convex nature of the problem, a convex
mixed-integer second-order cone relaxation is introduced. The proposed
convex relaxation is based on four key ideas: (1) the introduction of
variables for modeling the flux directions; (2) exact McCormick
relaxations; (3) on/off constraints; and (4) valid integer
cuts. Experimental results on the Belgian gas network and a test bed
of large-scale synthetic instances demonstrate three key findings:
\begin{enumerate}
\item The convex relaxation produces tight lower bounds with high
  computational efficiency;

\item The solution to the convex relxation can almost always be used
  to derive high-quality solutions to the original problem, leading to
  provably tight optimality gaps and, in some cases, global optimal
  solutions.

\item The proposed approach scales to large-scale instances.
\end{enumerate}

\noindent
The rest of this paper is organized as follows. Section
\ref{sec:Literature} presents the literature review. Section
\ref{sec:Brackground} introduces the problem formulation.  Section
\ref{section-relaxation} specifies the convex relaxation. Section
\ref{Sec:ComputationalExp} presents the computational results and
Section \ref{Sec:Conclusions} concludes the paper.

\section{Literature review} 
\label{sec:Literature}

The last four decades have seen an interest in natural gas planning
problems such as optimal design, optimal reinforcement, and optimal
expansion of gas pipeline systems.  Algorithms for these problems can
be classified in a number of different ways such as exact approaches
\citep{Ref_AndreEtAl2009,Ref_BonnansEtAl2011,Ref_EdgarEtAl1978,Ref_EdgarAndHimmelblau2001,Ref_DeWolf2004}
and heuristics
\citep{Ref_Andre2010,Ref_BoydEtAl1994,Ref_HumpolaEtAl2015b,Ref_AndreEtAl2009,Ref_HumpolaAndFugenschuh2014a}.
Exact methods include cutting planes
\citep{Ref_Atamturk2002,Ref_HumpolaAndFugenschuh2014b,Ref_HumpolaEtAl2015,Ref_Poss2011}
and branch-and-bound
\citep{Ref_Andre2010,Ref_ElshiekhEtAl2013,Ref_HumpolaAndFugenschuh2015}
and they use a variety of commercial
\citep{Ref_BakhouyaAndDeWolf2008,Ref_ElshiekhEtAl2013,Ref_SolimanAndMurtagh1982}
and open-source
\citep{Ref_PfetschEtAl2012,Ref_UsterAndDilaveroglu2014} solvers.  Like
this paper, much of the literature relies on approximations and
relaxations to improve the tractablity of the underlying planning
problems.  Examples include continuous relaxations of the discrete
design variables
\citep{Ref_DeWolfAndSmeers1996,Ref_HansenEtAl1991,Ref_SolimanAndMurtagh1982}
and approximation or relaxations of constraints
\citep{Ref_BabonneauEtAl2012,Ref_BakhouyaAndDeWolf2008,Ref_HumpolaAndFugenschuh2015,Ref_Poss2011}. Common
approaches for implementing these approximations/relaxations include
succesive linear programming
\citep{Ref_DeWolfEtAl1991,Ref_HansenEtAl1991,Ref_ONeillEtAl1979,Ref_WilsonEtAl1988}
and piecewise linearizations
\citep{Ref_CorreaPosadaAndSanchezMartin2014,Ref_MarkowitzAndManne1957,Ref_Vajda1964,Ref_ZhengEtAl2010}.

{\em The contribution of this paper is a novel Second-Order Cone (SOC)
  relaxation that efficiently addresses the design of large-scale
  cyclic networks for which flow directions are unknown.}  The model
captures physical, operational, contractual, and on/off constraints
and includes models of regular pipelines, valves, short pipes, control
valves, compressor stations, and regulators. Its dual solutions can
almost always be converted to high-quality or optimal primal
solutions.  To the best of our knowledge, this combination of features
has not appeared in the literature. Our paper focuses on the cost of
building the network but can be generalized to include operational
costs as well.

We now provide an in-depth review of the most relevant works in the
area of natural gas expansion planning problems. One of the earliest
papers that addresses natural gas design problems is
\citep{Ref_EdgarEtAl1978}. {\em It focuses on the optimal design of
  gunbarrel and tree-shaped networks.} Their objective minimizes the
yearly cumulative operational and investment costs. The optimization
variables include pipeline diameters, compression ratios, and the
number of compressors. In their later work,
\cite{Ref_EdgarAndHimmelblau2001} present a MINLP formulation for the
optimal design of a gas transmission network where the number of
compressor stations, the length and diameter of the pipeline sections,
and the inlet and outlet pressures at each stations are
optimized. They solve a simplified version of the problem in
GAMS~\citep{refGAMS} for a small instance \citep{Ref_EdgarEtAl2001}.

\cite{Ref_HansenEtAl1991} and \cite{Ref_SolimanAndMurtagh1982} propose
a continuous relaxation for the network design problem. While
\cite{Ref_HansenEtAl1991} apply a successive linear programming method
where a linear subproblem is solved to adjust the discrete choice of
diameters, \cite{Ref_SolimanAndMurtagh1982} apply the commercial NLP
solver MINOS~\cite{Ref_MurtaghAndSaunders1998} to handle the relaxed
subproblem.  \cite{Ref_ONeillEtAl1979} and ~\cite{Ref_WilsonEtAl1988}
focus on a problem where integer variables are used for the
operational state of compressor stations and they also implement a
method based on successive linear programming to solve the problem.

\cite{Ref_DeWolfAndSmeers1996} address the optimal dimensioning of a
known pipe network topology with an objective that combines the cost
of purchasing gas and the capital expenditures for expansion. The
authors formulate the problem as a continuous NLP that selects
pipeline diameters and solves the problem by means of a local
optimizer. Based on this problem, \cite{Ref_DeWolf2004} derives
conditions under which this problem is convex. Through the use of
variational inequality theory, they show convexity of the nonlinear
gas flow system under the assumption that the gas net inlet (pressure)
is fixed at all supply and demand
nodes. ~\cite{Ref_BakhouyaAndDeWolf2008} also present a case study on
the same problem with separable transportation and gas objectives that
leads to a two-stage problem formulation. In addition to design
variables for the optimal pipe diameters, the authors add investment
variables representing the maximal power of compressor stations to
balance the pipeline construction costs and capital expenditures for
increasing power in the compressor units. The authors find an initial
solution by solving a convex problem where all pressure constraints
are relaxed. Then, the complete problem is locally solved by means of
the GAMS/CONOPT solver. In these works, numerical experiments are
primarily focused on the Belgian gas transmission network.

\cite{Ref_AndreEtAl2009} present a MINLP model to solve the investment
cost minimization problem for an existing gas system that includes
pipelines and regulators and omits compressor stations. The goal is to
identify a set of pipeline sections to reinforce and to select an
optimal diameter size for these sections based on a discrete set of
diameters. Under the assumption that the network is radial (the head
loss equations are convex when flows are fixed), the authors propose a
continuous relaxation of the pipe diameters (continuous intervals).  A
branch-and-bound approach for a unique maximal demand scenario is
applied to a segment of the French high-pressure natural gas
transmission system. A complete review and extensions of these
findings are provided in~\citep{Ref_Andre2010}.

\cite{Ref_BabonneauEtAl2009,Ref_BabonneauEtAl2012} focus on the design
and operation of a natural gas transmission system while minimizing
investment, purchase, and transportation costs. The authors propose an
approach based on a minimum energy principle that transforms the
non-linear non-convex optimization problem into a convex problem. The
underlying convex, bi-objective formulation is an approximation of the
investment cost function and the cost of energy to transport the gas.
Their continuous formulation is applied to non-divisible constraints
such as a limited number of available commercial pipe dimensions.

\cite{Ref_BonnansEtAl2011} presents several problems that include the
minimization of compressor ratios and the sum of operations and
investment costs. The authors propose a global optimization technique
that is based on the combination of interval analysis with constraint
propagation.

\cite{Ref_ZhengEtAl2010} discusses different optimization models in the
natural gas industry, including the compressor station allocation problem,
the least gas purchase problem and optimal dimensioning of gas pipelines. The
authors review solution techniques to solve the underlying models which
include a piecewise linear programming algorithm and a branch-and-bound
algorithm.

\cite{Ref_ElshiekhEtAl2013} presents a model to optimize the design
and operation of the Egyptian gas system, where continuous design
variables for the length and diameter of pipelines are considered
along with a modified Panhandle
equation~\citep{Ref_CoelhoAndPinho2007}. The complete model is directly
solved by means of the computer-aided optimization software
LINGO~\citep{Ref_LINDO1997}.

\cite{Ref_UsterAndDilaveroglu2014} address the cost minimization
problem of designing a new natural gas transmission system and
expanding an existing gas system. The authors propose a mathematical
formulation to tackle the design/expansion network problem for a given
multi-period planning horizon. The underlying MINLP model is
formulated in AMPL and solved approximately with
Bonmin~\citep{Ref_BonamiAndLee2013}.

\cite{Ref_HumpolaAndFugenschuh2014b} and \cite{Ref_HumpolaEtAl2015}
present valid inequalities for a MINLP model of a design problem in
gas transmission systems. Different relaxations are applied to the
subproblems created after branching on the additive and design
variables for the active and passive components. The resulting passive
transmission subproblems, which are referred to as leaf problems,
admit slack variables to independently relax the pressure domains and
the flow conservation constraints. The proposed cutting planes aim at
reducing the CPU time of a branch-and-cut-based outer approximation
applied to the full model where construction costs are defined by a
global constant. \cite{Ref_Atamturk2002} and \cite{Ref_Poss2011} also
propose valid inequalities to reinforce the relaxation approach to the
network design structure.

\cite{Ref_HumpolaAndFugenschuh2015} examines different (convex)
relaxations for subproblems created while applying a branch-and-bound
technique to a nonlinear network design problem. Cutting planes
on the nonlinear potential loss
constraints are used to strengthen the relaxed subproblems.

\cite{Ref_PfetschEtAl2012} focuses on the validation of nomination
problem while considering regular pipes and valves, control valves,
compressors and regulators. The authors describe a two-stage approach
to solve the resulting MINLP problem and propose several modeling
techniques and approaches to account for, e.g., pressure losses. They
also developed several large test cases \citep{Ref_GasLib2014}. These
problems form the basis for many of the problems we consider in this
paper.

\section{Problem Formulation} 
\label{sec:Brackground}

This section derives the problem formulation (as a disjunctive
program) in stepwise refinements. It starts by deriving a disjunctive
formulation that is then refined by introducing flux variables.

\subsection{The Disjunctive Formultion}

Gas dynamics along a pipe is
described by a set of partial differential equation (PDE) with both
spatial and temporal dimensions
\citep{osiadacz1987simulation,87TT,05Sar}:

\begin{eqnarray}
&& \partial_t\rho+\partial_x (\rho v)=0,\label{density_eq}\\
&& \partial_t (\rho v)+\partial_x (\rho v^2)+\partial_x p=- \frac{\bm f}{2\bm D}{\rho v |v|}- {\bm g }\sin {\bm \alpha} \rho,\label{momenta_eq}\\
&& p=\bm {ZRT}\rho .\label{thermodynamic_eq}
\label{state_eq}
\end{eqnarray}
Gas velocity $v$, pressure $p$, and density $\rho$ are defined for
every point $x$ along the pipe and evolve over time $t$.  $\bm Z$
represents the gas compressibility factor, $\bm T$ the temperature,
and $\bm R$ the gas constant.

Equation \eqref{density_eq} enforces mass conservation, Equation
\eqref{momenta_eq} describes momentum balance, and Equation
\eqref{state_eq} defines the ideal gas thermodynamic relation. In
Equation \eqref{momenta_eq}, the first term on the right-hand side
represents the friction losses in a pipe of diameter $\bm D$ with
friction factor $\bm f$.  The second term accounts for the gain or
loss of momentum due to gravity $\bm g$ if the pipe is tilted by an
angle $\bm \alpha$.  In practice, frictional losses dominate the
gravitational term which is dropped. One can also safely assume that
the temperature does not fluctuate significantly along a pipe. If
temperature gradients are significant, a spatial decomposition,
splitting the pipe into temperature stable segments, can be adopted.

Taking into account these assumptions, Equations
\eqref{density_eq},\eqref{momenta_eq}, and \eqref{thermodynamic_eq} are
rewritten in terms of pressure $p$ and mass flux $\phi=\rho v$:
\begin{eqnarray}
&& \partial_t p=-\bm{ZRT}\partial_x \phi,\label{density_eq1}\\
&& \partial_x p^2=- \frac{\bm {f Z R T}}{2\bm D} \phi |\phi|,\label{momenta_eq1}
\end{eqnarray}
In this work, we assume that the system has reached a steady state
after its first commissioning and hence all time derivatives are set
to zero. Given this assumption, a Graph Transmission Network (GTN) is
represented by a graph $\mathcal G = (\mathcal N,\mathcal A)$ where
$\mathcal N$ denotes the set of nodes representing connection points
and $\mathcal A$ denotes the set of arcs. An arc is a triplet
$(a,i,j)$ consisting of a unique identifier $a$ linking nodes $i$ and
$j$.  For convenience, such a triplet $(a,i,j)$ will be denoted by
$a_{ij}$ in the following. Observe that parallel arcs can link the
same pair of nodes, e.g., we have arcs $a_{ij}$ and $a^{*}_{ij}$ in a
GTN where $a$ and $a^{*}$ are the unique identifiers of these arcs.

By setting the time derivatives to zero, the total gas mass flux along
a pipe $a_{ij}$ becomes constant, i.e., $\phi_i = \phi_j = \phi_{a}$.
Hence Equations \eqref{density_eq1} and \eqref{momenta_eq1} simplify
to
\begin{equation}
p_i^2-p_j^2=\bm{w}_{a} \phi_{a} |\phi_{a}|.\label{eq:weymouth}
\end{equation}
where $\bm{w_{a}} = \frac{\bm{Z R}\bm{f} _{a}\bm{T}_{a}}{{2\bm D_{a}}}$.


\paragraph{Gas System Components}

The problem formulation considers \emph{pipes}, \emph{compressors},
\emph{short pipes}, \emph{resistors}, and \emph{valves}.  Compressors,
short pipes, and valves are modelled as lossless pipelines, i.e.,
$\bm{w_{a}} = 0$.  A compressor installed on arc $a_{ij}$ can
increase/decrease the pressure ratio $\alpha_{a} = {p_j}/{p_i}$,
within the bounds ${\bm \alpha}^l_{a}$ and ${\bm \alpha}^u_a$, where
${\bm \alpha}^l_{a}=1$ and ${\bm \alpha}^u_a \ge 1$ is typical for
most compressors.  A \emph{bi-directional compressor} can perform
compression based on the flux direction, i.e., it is able to invert
the ratio to $\alpha_{a} = {p_i}/{p_j}$ if the flux is going from $j$
to $i$.  A standard valve features a binary on/off switch and a
\emph{control valve} has a continuous switching mechanism to adjust
pressure.  Thus a valve installed on arc $a_{ij}$ can
increase/decrease the pressure ratio $\alpha_{a} = {p_j}/{p_i}$,
within the bounds ${\bm \alpha}^l_{a}$ and ${\bm \alpha}^u_a$, where
${\bm \alpha}^l_{a}> 0$ and ${\bm \alpha}^u_a \le 1$ is typical for
most control valves and ${\bm \alpha}^l_{a} = {\bm \alpha}^u_a = 1$
for all valves.  Finally, a resistor is modelled as a pipeline with a
particular (small) loss coefficient ($\bm{w}$).

\paragraph{Expansion Variables}

The set of arcs $\mathcal A = {\cal A}_{e} \cup {\cal A}_{n}$ includes
existing arcs ${\cal A}_{e} = {\cal P}_{e} ~\cup~ {\cal C}_{e} ~\cup~
{\cal V}_{e}$, as well as new arcs ${\cal A}_{n} = {\cal P}_{n} ~\cup~
{\cal C}_{n}$.  In this notation, ${\cal P}_{e}$ denotes the set of
installed pipelines, resistors, and short pipes. ${\cal C}_{e}$ and
${\cal V}_{e}$ denote the set of existing compressors and valves
(control and regular) respectively.  ${\cal P}_{n}$ and ${\cal C}_{n}$
denote the set of new pipelines and new compressors respectively.  A
binary variable $z^{p}_{a}$ is assigned for each new pipe $a_{ij}$ in
${\cal P}_{n}$ to model the expansion decision, i.e., $z^{p}_{a}=1$ if
pipeline $a_{ij}$ is installed and $z^{p}_{a}=0$ otherwise.  Variables
$z^{c}_{ij}$ $((i,j) \in {\cal C}_{n})$ have an equivalent
interpretation for new compressors.  A binary variable $v_a$ is used
to control the switching actions of valves.

\paragraph{Disjunctive Formulation}

\begin{figure}[t]
  \centering
  \includegraphics[width=0.5\columnwidth]{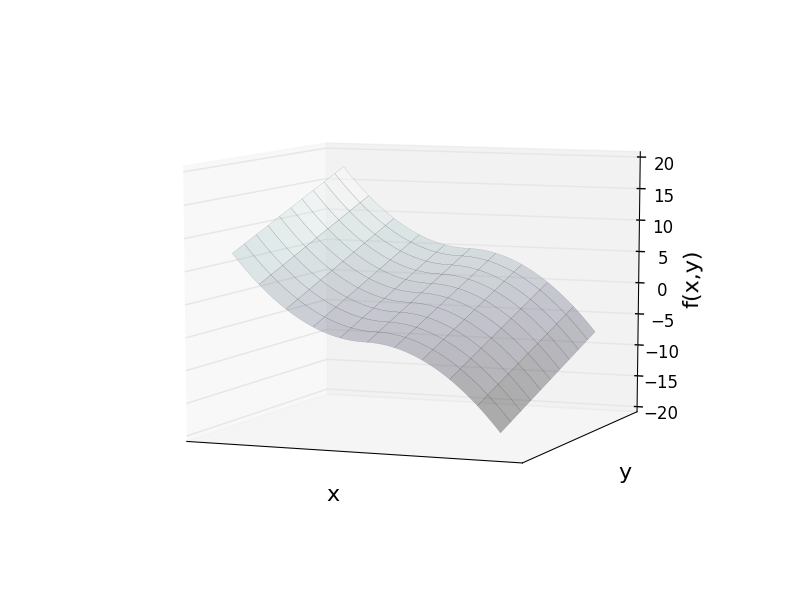}
  \caption{The Gas Flow Equation $f(x,y) = y - {\bm w}x|x|$.} 
\label{fig:weymouth}
\end{figure}

Since the pressure variables only appear in a square form, the
formulation uses the variable substitution $\beta_i = p_i^2$ $(i \in
\mathcal N)$. Equations \eqref{eq:weymouth} can be written as
\begin{equation}
  \beta_i-\beta_j=\bm{w}_{a} \phi_{a} |\phi_{a}| \;\;\; (a_{ij} \in {\cal P}_e) \label{eq:drop_o}
\end{equation}
Figure \ref{fig:weymouth} illustrates the curve of the function
$f(x,y) = y - {\bm w}x|x|$ defined by the pressure drop equation
\eqref{eq:drop_o}.

Since bi-directional compressor constraints depend on the flux
direction, they can only be modelled using on/off or disjunctive
constraints \citep{Ref_HijaziEtAl2010,Ref_HijaziEtAl2012}. i.e.,
\begin{equation} 
\label{eq:comp}
\begin{cases}
\beta_i {\bm \alpha}^l_{a} \le \beta_j \le \beta_i {\bm \alpha}^u_{a},~\text{if } \phi_{a}\ge0\\
\beta_j {\bm \alpha}^l_{a} \le \beta_i \le \beta_j {\bm \alpha}^u_{a},~\text{if } \phi_{a}\le0.\\
\end{cases}
\end{equation}

\noindent
Given a set of injection (resp. demand) nodes \mbox{${\cal
    I}~(\text{resp. } {\cal D}) \subseteq \mathcal N$} with mass flux
injection/demand $\bm{q}_i$, the problem consists in finding an
assignment of the expansion variables $z^{p}_{a} \;(a_{ij} \in {\cal
  P}_{n})$, node pressures $p_i \; (i\in{\cal N})$, and edge flows
$\phi_{a} \; (a_{ij}\in{\cal A})$, satisfying the Weymouth equations
\eqref{eq:drop_o}, the compressor constraints \eqref{eq:comp}, and
the following node conservation constraints:
\begin{equation*}
  \sum_{a_{ij}\in{\cal A}}\phi_{a} = \sum_{a_{ji}\in{\cal A}}\phi_{a} + \bm{q}_i \; (i\in{\cal N})
\end{equation*}
where $q_i = 0$ for all $i \in N \setminus ({\cal I} \cup {\cal D})$.
Note that, in the steady-state model, injections are balanced, i.e.,
$\sum_{i\in{\cal N}} \bm{q}_i=0$.  The objective is to minimize the
cost of expansion:
\begin{equation*}
\min \sum_{a_{ij}\in{\cal P}_n}{\bm c}_{a}z^{p}_{a} + \sum_{a_{ij} \in {\cal C}_n}{\bm c}_{a}z^{c}_a
\end{equation*}
where ${\bm c}_{a}$ represents the cost of installing a new
pipeline. The disjunctive formulation of the problem incorporating
these ideas is presented in Model \ref{mod:disj}, where $\bm
{\beta^{l}}_i = (\bm {\alpha^{l}}_i)^2$ and $\bm {\beta^{u}}_i = (\bm
{\alpha^{u}}_i)^2$. 

\begin{model}[t]
\caption{The Disjunctive Formulation of the GTNEP.}
\vspace{-0.3cm} 
\label{mod:disj}
\begin{subequations}
\begin{align}
\mbox{\bf variables:} \nonumber \\
& \beta_i \in [\bm {\beta^{l}}_i , \bm {\beta^{u}}_i]\;\; \forall i\in \mathcal N \mbox{ - squared pressure level variables} \nonumber \\
& \phi_{a} \in \rit \;\; \forall a_{ij}\in{\cal A} \mbox{ - mass flux on pipe (i,j)} \nonumber \\
& z^p_{a} \in \{0,1\} \;\; \forall a_{ij} \in {\cal P}_n \mbox{ - binary expansion variables for pipes} \nonumber \\
& z^c_{a} \in \{0,1\} \;\; \forall a_{ij} \in {\cal C}_n \mbox{ - binary expansion variables for compressors} \nonumber \\
& v_{a} \in \{0,1\} \;\; \forall a_{ij} \in {\cal CV}_e \cup {\cal V}_e \mbox{ - binary switch variables for valves} \nonumber \\
\mbox{\bf objective:} \nonumber \\
&\min \sum_{a_{ij}\in{\cal P}_n}{\bm c}_{a}z^{p}_{a} + \sum_{a_{ij} \in {\cal C}_n}{\bm c}_{a}z^{c}_{a}\\
\mbox{\bf subject to:} \nonumber \\
& \sum_{a_{ij}\in{\cal A}}\phi_{a} = \sum_{a_{ji}\in{\cal A}}\phi_{a} + \bm{q}_i ,~\forall i\in{\cal N} \\
&\beta_i-\beta_j=\bm{w}_{a} \phi_{a} |\phi_{a}|,~\forall a_{ij} \in {\cal P}_e\\
&z^p_{a}(\beta_i-\beta_j)=\bm{w}_{a} \phi_{a} |\phi_{a}|,~\forall a_{ij} \in {\cal P}_n\\
&\beta_i {\bm \alpha}^l_{a} \le \beta_j \le \beta_i {\bm \alpha}^u_{a},~\text{if } \phi_{a}\ge0,~ \forall a_{ij}\in{\cal C}_e,\\
&\beta_j {\bm \alpha}^l_{a} \le \beta_i \le \beta_j {\bm \alpha}^u_{a},~\text{if } \phi_{a}\le0,~ \forall a_{ij}\in{\cal C}_e,\\
&\beta_i {\bm \alpha}^l_{a} \le \beta_j \le \beta_i {\bm \alpha}^u_{a},~\text{if } \phi_{a} \ge 0  \text{ and } z^{c}_{a} = 1,~ \forall a_{ij}\in{\cal C}_n,\\
&\beta_j {\bm \alpha}^l_{a} \le \beta_i \le \beta_j {\bm \alpha}^u_{a},~\text{if } \phi_{a} \le 0  \text{ and } z^{c}_{a} = 1,~ \forall a_{ij}\in{\cal C}_n,\\
&\phi_{a} = 0  \text{ if } z^{c}_{a} = 0,~ \forall a_{ij}\in{\cal C}_n \\
&\beta_i {\bm \alpha}^l_{a} \le \beta_j \le \beta_i {\bm \alpha}^u_{a},~\text{if } \phi_{a} \ge 0  \text{ and } v^{c}_{a} = 1,~ \forall a_{ij}\in{\cal V}_e,\\
&\beta_j {\bm \alpha}^l_{a} \le \beta_i \le \beta_j {\bm \alpha}^u_{a},~\text{if } \phi_{a} \le 0  \text{ and } v^{c}_{a} = 1,~ \forall a_{ij}\in{\cal V}_e,\\
&\phi_{a} = 0  \text{ if } v^{c}_{a} = 0,~ \forall a_{ij}\in{\cal C}_e
%
\end{align}
\end{subequations}
\end{model}

\subsection{The Formulation Based on Flux Direction Variables}

This section presents a second formulation using flux direction
variables to account for the disjunctive nature of the
constraints. For every arc $a_{ij} \in {\mathcal{A}}$, we introduce
two binary variables $y^{+}_{a}$ and $y^{-}_{a} \in \{0,1\}$ with the
following semantics: $y^{+}_{a} = 1$ (resp. $y^{-}_{a} = 1$) if the
flux moves from $i$ to $j$ (resp. from $j$ to $i$) and 0 otherwise.
The mass flux direction is captured by the following system of
constraints:
\[
\begin{cases}
&(1-y^+_{a})\sum\limits_{k\in{\cal I}} \bm{q}_k\le \phi_{a} \le (1-y^-_{a})\sum\limits_{k\in{\cal I}} \bm{q}_k,\\[0.5em]
&\left(1 - y^+_{a}\right)\bm {\beta^{l}}_i \le \beta_{i} - \beta_{j} \le \left(1 - y^-_{a}\right)\bm {\beta^{u}}_i,\\[0.5em]			
&y^+_{a} + y^-_{a} = 1.\\
\end{cases}
\]
The first constraint ensures that $y^+_{a} = 1$ (resp. $y^-_{a} = 1$)
if and only if $\phi_a \geq 0$ (resp. $\phi_a \leq 0$).  Note that
$\sum\limits_{k\in{\cal I}} \bm{q}_k$ is an upper bound to the mass
flux in a pipe. The second constraint enforces a similar condition
for the pressure difference.  Using the variables and constraints, the
pressure drop equation can now be written without absolute value as
\[
\left(y^{+}_{a} - y^{-}_{a}\right)(\beta_{i} - \beta_{j}) = \bm{w}_{a} \phi_{a}^2
\]
and the bi-directional compressor constraints are written as 
\begin{align}
& \beta_i {\bm \alpha}^l_{a} - (1 - y^+_{a})(\bm {\beta^{u}}_i{\bm \alpha}^l_{a} - \bm {\beta^{l}}_j) \le \beta_j \le \beta_i {\bm \alpha}^u_{a} +  (1 - y^+_{a})(\bm {\beta^{u}}_j - \bm {\beta^{l}}_i{\bm \alpha}^u_{a}),~\forall a_{ij} \in {\cal C}_e \label{eq:MINLP:compressor}\\
& \beta_j {\bm \alpha}^l_{a} - (1 - y^-_{a})(\bm {\beta^{u}}_j{\bm \alpha}^l_{a} - \bm {\beta^{l}}_i) \le \beta_i \le \beta_j {\bm \alpha}^u_{a} +  (1 - y^-_{a})(\bm {\beta^{u}}_i - \bm {\beta^{l}}_j{\bm \alpha}^u_{a}),~\forall a_{ij} \in {\cal C}_e\\
& \beta_i {\bm \alpha}^l_{a} - (2 - y^+_{a} - z^c_a)(\bm {\beta^{u}}_i{\bm \alpha}^l_{a} - \bm {\beta^{l}}_j) \le \beta_j \le \beta_i {\bm \alpha}^u_{a} +  (2 - y^+_{a} - z^c_a)(\bm {\beta^{u}}_j - \bm {\beta^{l}}_i{\bm \alpha}^u_{a}),~\forall a_{ij} \in {\cal C}_n\\
& \beta_j {\bm \alpha}^l_{a} - (2 - y^-_{a} - z^c_a)(\bm {\beta^{u}}_j{\bm \alpha}^l_{a} - \bm {\beta^{l}}_i) \le \beta_i \le \beta_j {\bm \alpha}^u_{a} +  (2 - y^-_{a} - z^c_a)(\bm {\beta^{u}}_i - \bm {\beta^{l}}_j{\bm \alpha}^u_{a}),~\forall a_{ij} \in {\cal C}_n \label{eq:MINLP:compressorend} \\
&-z_a^c \sum_{i \in \cal{I}} \le \phi_{a} \le z_a^c \sum_{i \in \cal{I}} ,~\forall a_{ij} \in {\cal C}_n  \label{eqn:compflow}
\end{align}

\noindent
The bi-directional valve constraints are written as
\begin{align}
& \beta_i {\bm \alpha}^l_{a} - (2 - y^+_{a} - v_a)(\bm {\beta^{u}}_i{\bm \alpha}^l_{a} - \bm {\beta^{l}}_j) \le \beta_j \le \beta_i {\bm \alpha}^u_{a} +  (2 - y^+_{a} - v_a)(\bm {\beta^{u}}_j - \bm {\beta^{l}}_i{\bm \alpha}^u_{a}),~\forall a_{ij} \in {\cal V}_e\\
& \beta_j {\bm \alpha}^l_{a} - (2 - y^-_{a} - v_a)(\bm {\beta^{u}}_j{\bm \alpha}^l_{a} - \bm {\beta^{l}}_i) \le \beta_i \le \beta_j {\bm \alpha}^u_{a} +  (2 - y^-_{a} - v_a)(\bm {\beta^{u}}_i - \bm {\beta^{l}}_j{\bm \alpha}^u_{a}),~\forall a_{ij} \in {\cal V}_e \label{eq:MINLP:valveend} \\
&-v_a \sum_{i \in \cal{I}} \le \phi_{a} \le v_a \sum_{i \in \cal{I}} ,~\forall a_{ij} \in {\cal V}_e  \label{eq:MINLP:valveflow}
\end{align}

\noindent
The complete Mixed-Integer NonLinear Programming (MINLP) formulation
based on flux direction variables is summarized in Model
\ref{mod:MINLP}. The continuous relaxation of Model \ref{mod:MINLP} is
non-convex due to Constraints \eqref{eq:drop}-\eqref{eq:drop_}.

\begin{model}[t]
\caption{The MINLP Formulation of the GTNEP.}
\vspace{-0.3cm} 
\label{mod:MINLP}
\begin{subequations}
\begin{align}
\mbox{\bf variables:} \nonumber \\
& \beta_i \in [\bm {\beta^{l}}_i , \bm {\beta^{u}}_i]\;\; \forall i\in \mathcal N \mbox{ - squared pressure level variables} \nonumber \\
& \phi_{a} \in \rit \;\; \forall a_{ij}\in{\cal A} \mbox{ - mass flux on pipe (i,j)} \nonumber \\
& z^p_{a} \in \{0,1\} \;\; \forall a_{ij} \in {\cal P}_n \mbox{ - binary expansion variables for pipes} \nonumber \\
& y^+_{a}, y^-_{a} \in \{0,1\},~\forall a_{ij} \in {\cal A} \mbox{ - binary flux direction variables }\nonumber \\
& z^c_{a} \in \{0,1\} \;\; \forall a_{ij} \in {\cal C}_n \mbox{ - binary expansion variables for compressors} \nonumber \\
& v_{a} \in \{0,1\} \;\; \forall a_{ij} \in {\cal CV}_e \cup {\cal V}_e \mbox{ - binary switch variables for valves} \nonumber \\
\mbox{\bf objective:} \nonumber \\
&\min \sum_{a_{ij}\in{\cal P}_n}{\bm c}_{a}z^{p}_{a} + \sum_{a_{ij} \in {\cal C}_n}{\bm c}_{ij}z^{c}_{ij}\\
\mbox{\bf subject to:} \nonumber \\
& \sum_{a_{ij}\in{\cal A}}\phi_{a} = \sum_{a_{ji}\in{\cal A}}\phi_{a} + \bm{q}_i ,~\forall i\in{\cal N} \\
&\left(y^{+}_{a} - y^{-}_{a}\right)(\beta_{i} - \beta_{j}) = \bm{w}_{a} \phi_{a}^2, ~\forall a_{ij} \in {\cal P}_e\label{eq:MINLP:drop}\\
&z^p_{a}\left(y^{+}_{a} - y^{-}_{a}\right)(\beta_{i} - \beta_{j}) = \bm{w}_{a} \phi_{a}^2, ~\forall a_{ij} \in {\cal P}_n\label{eq:MINLP:drop_}\\
&\left(1-y^+_{a}\right)\sum_{k\in{\cal I}} \bm{q}_k \le \phi_{a} \le \left(1-y^-_{a}\right)\sum_{k\in{\cal I}} \bm{q}_k, ~\forall a_{ij} \in {\cal A}\\
&\left(1 - y^+_{a}\right)\bm {\beta^{l}}_i \le \beta_{i} - \beta_{j} \le \left(1 - y^-_{a}\right)\bm {\beta^{u}}_i,~\forall a_{ij} \in {\cal P}\\						
& (\ref{eq:MINLP:compressor}-\ref{eq:MINLP:valveflow}) \\
&y^+_{a} + y^-_{a} = 1,~\forall a_{ij} \in {\cal A} 
%
\end{align}
\end{subequations}
\end{model}

\section{A Convex Relaxation of the GPNEP}
\label{section-relaxation}

This section introduces a new mixed-integer second-order cone
relaxation for Model~\ref{mod:MINLP}.

\subsection{The Variables}

For every pipe $a_{ij} \in {\cal P}$, the relaxation introduces the
auxiliary variable $\gamma_{a}$ representing the product in Equations
\eqref{eq:MINLP:drop}-\eqref{eq:MINLP:drop_}, i.e.,
\begin{equation}
\gamma_{a} = \left(y^{+}_{a} - y^{-}_{a}\right)(\beta_{i} - \beta_{j}), ~\forall a_{ij} \in {\cal P}, \label{eq:bilinear}
\end{equation}
This product is then linearlized by a standard relaxation introduced by \cite{Ref_McCormick1976} for bilinear functions, i.e.,
\begin{align}
&{\gamma}_{a} \ge \beta_j - \beta_i + \left(\bm {\beta}^l_i - \bm {\beta}^u_j\right) (y^{+}_{a} - y^{-}_{a} + 1)\label{eq:mc1}\\
&{\gamma}_{a} \ge \beta_i - \beta_j + \left(\bm {\beta}^u_i - \bm {\beta}^l_j\right) (y^{+}_{a} - y^{-}_{a} - 1)\label{eq:mc2}\\
&{\gamma}_{a} \le \beta_j - \beta_i + \left(\bm {\beta}^u_i - \bm {\beta}^l_j\right) (y^{+}_{a} - y^{-}_{a} +1)\label{eq:mc3}\\
&{\gamma}_{a} \le \beta_i - \beta_j +  \left(\bm {\beta}^l_i - \bm {\beta}^u_j\right) (y^{+}_{a} - y^{-}_{a} - 1)\label{eq:mc4}
\end{align}
This linearization is exact, since $\left(y^{+}_{a} - y^{-}_{a}\right)$
take only discrete values. 

\subsection{The Constraints}

The non-convex constraints \eqref{eq:MINLP:drop} can now be relaxed into 
\[
\gamma_{a} \ge \bm{w}_{a} \phi_{a}^2 \;\; (a_{ij} \in {\cal P}_e).
\]
The on/off constraints \eqref{eq:MINLP:drop_} represent another challenge for convexifying Model \ref{mod:MINLP}.
These constraints can be written as
\[
\gamma_{a} = \bm{w}_{a} \phi_{a}^2 \text{ if } z^p_{a}=1  \;\;(a_{ij} \in {\cal P}_n) 
\]
with a disjunctive second-order cone relaxation defined as
\[
\gamma_{a} \ge \bm{w}_{a} \phi_{a}^2, \text{ if } z^p_{a}=1 \;\; (a_{ij} \in {\cal P}_n).
\]
Perspective formulations introduced by \cite{Ref_HijaziEtAl2012} can
be used to formulate the convex hull of such on/off constraints,
giving the following rotated second-order cone constraint:
\[
z^p_{a}\gamma_{a} \ge \bm{w}_{a} \phi_{a}^2, ~\forall a_{ij} \in {\cal
  P}_n.
\]
The complete Mixed-Integer Second-Order Cone Programming (MISOCP)
relaxation is presented in Model \ref{mod:MISOCP}.

\begin{model}[t]
\caption{The MISOCP Relaxation for the GPNEP.}
\vspace{-0.3cm} 
\label{mod:MISOCP}
\begin{subequations}
\begin{align}
\mbox{\bf variables:} \nonumber \\
& \beta_i \in [\bm {\beta^{l}}_i , \bm {\beta^{u}}_i]\;\; \forall i\in \mathcal N \mbox{ - squared pressure level variables} \nonumber \\
& \phi_{a} \in \rit \;\; \forall a_{ij}\in{\cal A} \mbox{ - mass flux on pipe (i,j)} \nonumber \\
& z^p_{a} \in \{0,1\} \;\; \forall a_{ij} \in {\cal P}_n \mbox{ - binary expansion variables for pipes} \nonumber \\
& y^+_{a}, y^-_{a} \in \{0,1\},~\forall a_{ij} \in {\cal A} \mbox{ - binary flux direction variables }\nonumber \\
& \gamma_{a} \in \rit^+ \;\; \forall a_{ij} \in {\cal P} \mbox{ - auxiliary variables for bilinear terms} \nonumber \\
& z^c_{a} \in \{0,1\} \;\; \forall a_{ij} \in {\cal C}_n \mbox{ - binary expansion variables for compressors} \nonumber \\
& v_{a} \in \{0,1\} \;\; \forall a_{ij} \in {\cal CV}_e \cup {\cal V}_e \mbox{ - binary switch variables for valves} \nonumber \\
\mbox{\bf objective:} \nonumber \\
&\min \sum_{a_{ij}\in{\cal P}_n}{\bm c}_{a}z^{p}_{a} + \sum_{a_{ij} \in {\cal C}_n}{\bm c}_{ij}z^{c}_{ij}\\
\mbox{\bf subject to:} \nonumber \\
&(\ref{eq:mc1}-\ref{eq:mc4})\nonumber\\
& \sum_{a_{ij}\in{\cal A}}\phi_{a} = \sum_{a_{ji}\in{\cal A}}\phi_{a} + \bm{q}_i ,~\forall i\in{\cal N} \label{MISOCP:cons} \\
&\gamma_{a} \ge \bm{w}_{a} \phi_{a}^2, ~\forall a_{ij} \in {\cal P}_e\label{eq:drop}\\
&z^p_{a}\gamma_{a} \ge \bm{w}_{a} \phi_{a}^2, ~\forall a_{ij} \in {\cal P}_n\label{eq:drop_}\\
&-\left(1-y^+_{a}\right)\sum_{i\in{\cal I}} \bm{q}_i \le \phi_{a} \le \left(1-y^-_{a}\right)\sum_{i\in{\cal I}} \bm{q}_i, ~\forall a_{ij} \in {\cal A}\\
&\left(1 - y^+_{a}\right)\bm {\beta^{l}}_i \le \beta_{i} - \beta_{j} \le \left(1 - y^-_{a}\right)\bm {\beta^{u}}_i,~\forall a_{ij} \in {\cal P}\\	
& (\ref{eq:MINLP:compressor}-\ref{eq:MINLP:valveflow}) \\
&y^+_{a} + y^-_{a} = 1,~\forall a_{ij} \in {\cal A} 					
\end{align}
\end{subequations}
\end{model}

\subsection{The Integer Cuts}

The MINLP and MISOCP formulations presented in Models \ref{mod:MINLP}
and \ref{mod:MISOCP} can be strenghtened by introducing the following
valid integer cuts:

\begin{equation} 
\label{eq:injection}
\sum_{a_{ij} \in  \mathcal A} y^+_{a} + \sum_{a_{ji} \in \mathcal A} y^-_{a} \ge 1,~\forall i \in \mathcal{I}
\end{equation}

\begin{equation} 
\label{eq:demand}
\sum_{a_{ji} \in \mathcal A} y^+_{a} + \sum_{a_{ij} \in \mathcal A} y^-_{a} \ge 1,~\forall i \in \mathcal{D}
\end{equation}

\noindent
Constraints \eqref{eq:injection} are generated for each injection node
$i \in \mathcal I$: They state that at least one connected arc has an
outgoing flow, taking the orientation of the arc into account to
select the proper variables ($y^+_{a}$ for arcs leaving $i$ and
$y^-_{a}$ for arcs coming to $i$).  Constraints \eqref{eq:demand}
follow the same reasoning for demand nodes $i \in \mathcal D$.

For a node $i$ with degree two and no injection/demand ($\bm{q}_i=0$),
the following integer cut is valid
\begin{equation} 
\label{eq:degree2}
\begin{cases}
y^+_a = y^+_{a^*} ~\text{if } a_{ji},a^*_{ik} \in {\cal A} \\
y^+_a = y^-_{a^*} ~\text{if } a_{ji},a^*_{ki} \in {\cal A} \\
y^-_a = y^+_{a^*} ~\text{if } a_{ij},a^*_{ik} \in {\cal A} \\
y^-_a = y^-_{a^*} ~\text{if } a_{ij},a^*_{ki} \in {\cal A} \\
\end{cases}
\end{equation}
It can be easily derived using the flux conservation constraints
\eqref{MISOCP:cons} stating that, for a node with degree two and zero
injection/demand, the flux direction of the incoming arc determines
the flux direction of the outgoing arc.

Finally, we can derive integer cuts for parallel pipelines:
\begin{equation} \label{eq:parallel}
y^+_{a^{*}} = y^+_{a},~\forall a_{ij},a^*_{ij} \in \mathcal A.
\end{equation}
Equations \eqref{eq:parallel} state that parallel pipelines share the
same flow direction. The validity of this cut follows from the
pressure drop equations \eqref{eq:MINLP:drop} and the fact that
parallel pipelines share the same pair of pressure variables.

\subsection{Converting the Convex Relaxation in a Feasible Solution to  the GPNEP}
\label{section:conversion}

The solution to the relaxed Model \ref{mod:MISOCP} is not always
feasible for Model~\ref{mod:MINLP}. To obtain a feasible solution, we
fix all the binary variables and use a nonlinear optimization solver
to find a (locally) optimal solution to the resulting problem. When
the local solver does not converge to a feasible solution, we consider
primal solutions obtained when solving Model \ref{mod:MISOCP} and
repeat the process. 

\section{Computational Experiments} 
\label{Sec:ComputationalExp}

This section studies the performance of the proposed MINLP and MISOCP
models and compares them with a model using a piecewise linear
approximation. Section \ref{subsection:benchmarks} describes the
benchmarks and Section \ref{subsection:algorithms} the experimental
setting and the various algorithms used. Section
\ref{subsection:belgium} and \ref{subsection:scalability} report the
computational results on the Belgian network and larger networks
respectively, while Section \ref{subsection:integercuts} reports on
the importance of the integer cuts.

\subsection{The Benchmarks}
\label{subsection:benchmarks}

\subsubsection{The Belgian Network}

\begin{figure}[t]
\includegraphics[width=1\linewidth]{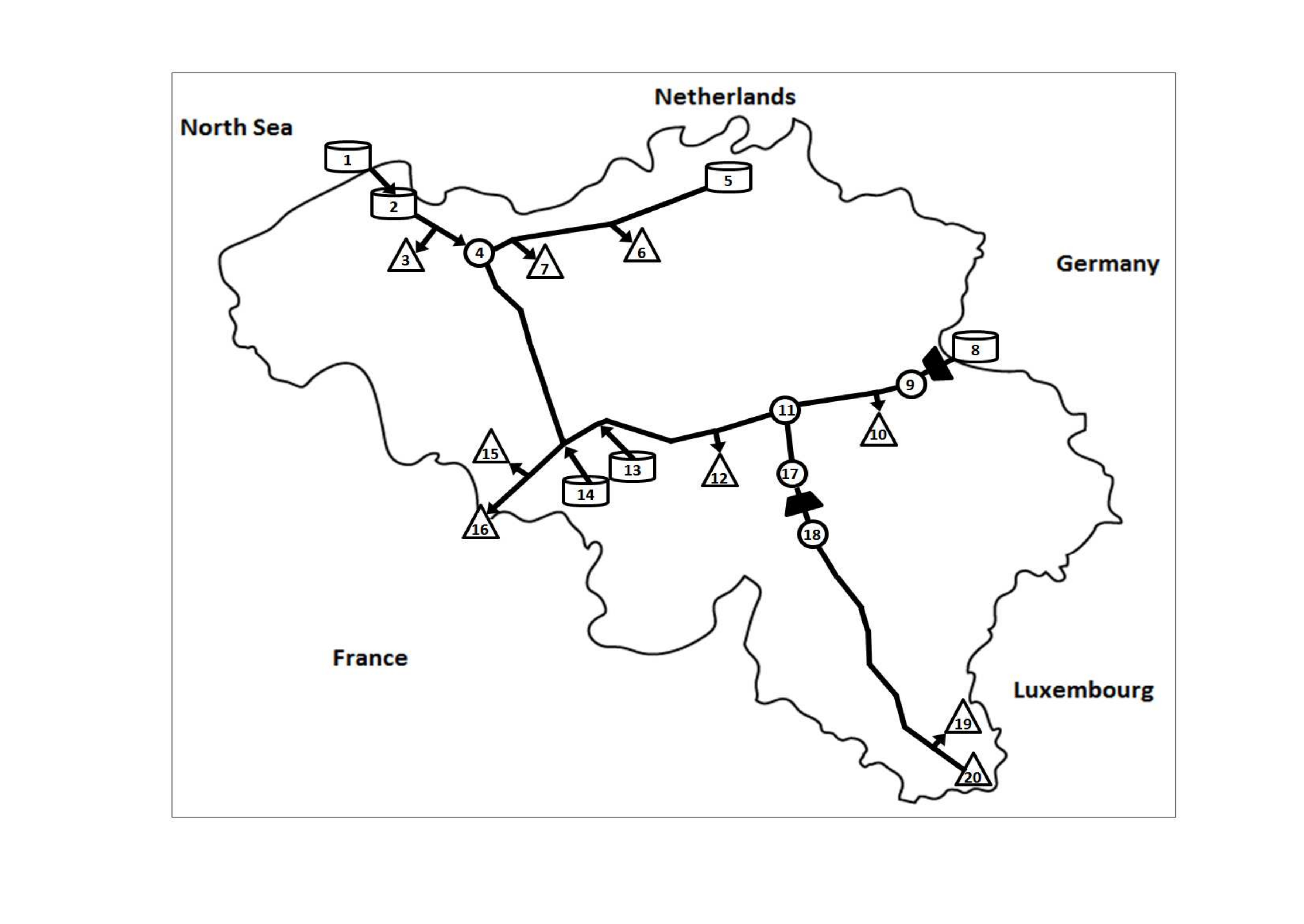}
\vspace{-1.5cm}
\caption{The Belgian Gas Network Base Configuration (case $A$).} 
\label{Fig:BelgianNetwork}
\end{figure}

\begin{figure*}[t]
\includegraphics[width=1\linewidth]{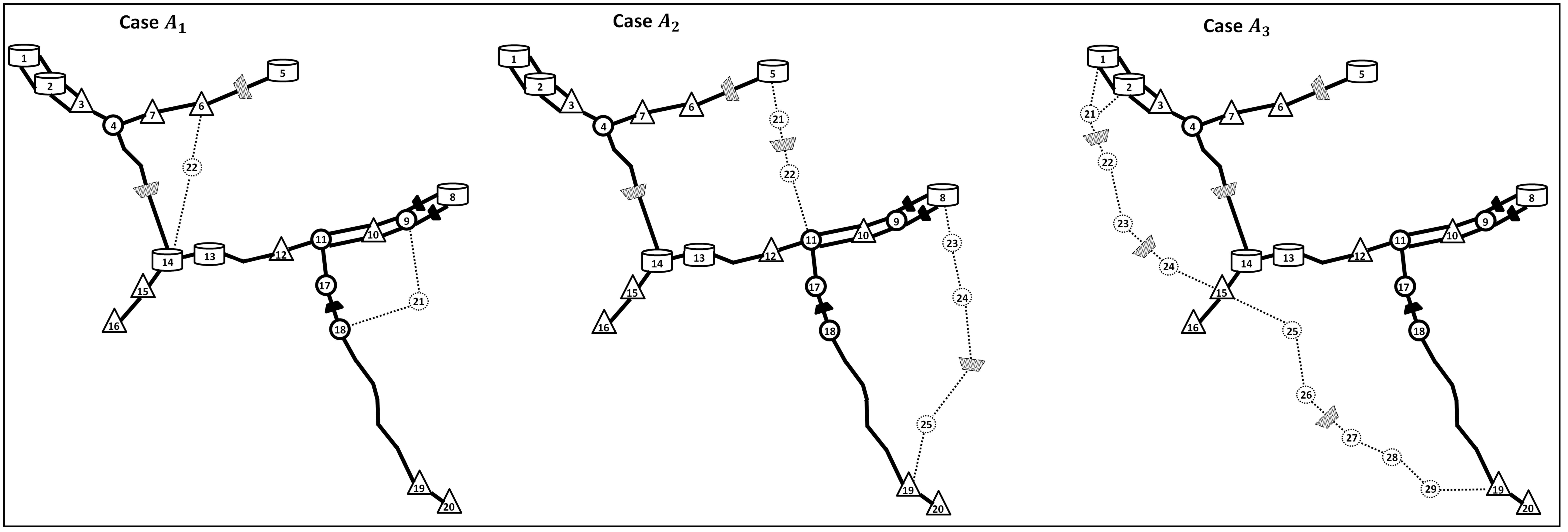}
\caption{The Belgian Gas Network Expansion Plans (Instances $A_1$--$A_3$).} 
\label{Fig:BelgianGasNetworkExpansions}
\end{figure*}

Table~\ref{Table:TestInstances} shows the list of test instances based
on the Belgian network depicted in Figure \ref{Fig:BelgianNetwork}.
The table shows, for each benchmark, the number of nodes, sources,
terminals, base pipelines, and compressor stations, as well as the
number of new components (pipelines and compressors) that can
potentially be added to the network topology. Note that benchmark $A$
in Table~\ref{Table:TestInstances} is the real Belgium gas
transmission network and Table~\ref{Table:BelgianNetwork} shows the
node characteristics for this 20-node, 24-pipeline, 3-compressor
network. The reader is referred to the appendix of
\citep{Ref_DeWolfAndSmeers2000} for further details on this network.
Instances $A_1-A_3$ captures various possible expansions to this base
network. Figure \ref{Fig:BelgianGasNetworkExpansions} and Tables
\ref{Table:NewNodes} and \ref{Table:NewPipes} depict the location of
the potential expansion plans and their associated data. The network
expansion plans were designed for the Belgian gas network in order to
capture events such as increase of the number of nominations and
forecasting demand at the city gates, as well as excessive stress of
the available supplies at the sources.

\begin{table}[t]
\centering
\begin{tabular}{clcccccrcc}
\hline\hline
    & & \multicolumn{8}{c}{Network configuration} \\ [0.5ex]
 \cline{3-10}
    & & & &  & \multicolumn{2}{c}{Base} & & \multicolumn{2}{c}{New} \\ \cline{6-7} \cline{9-10}
Ref & & $|{\cal N}|$ & $|{\cal I}|$ & $|{\cal D}|$ & $|{\cal P}_e|$ & $|{\cal C}_e|$ & & $|{\cal P}_n|$ & $|{\cal C}_n|$ \\
\cline{1-1} \cline{3-7} \cline{9-10}
 $A$   & & 20 & 6 & 9 & 24 & 3 & &  0 & 0 \\
 $A_1$ & & 22 & 6 & 9 & 24 & 3 & &  4 & 2 \\
 $A_2$ & & 25 & 6 & 9 & 24 & 3 & &  7 & 4 \\
 $A_3$ & & 29 & 6 & 9 & 24 & 3 & & 12 & 5 \\
 $B_1$ & & 20 & 6 & 9 & 0 & 0 & &  135 & 12 \\  
 $B_2$ & & 20 & 6 & 9 & 0 & 0 & &  135 & 12 \\  
 $B_3$ & & 20 & 6 & 9 & 0 & 0 & &  135 & 12 \\  
 $B_4$ & & 20 & 6 & 9 & 0 & 0 & &  135 & 12 \\  
 \hline

\end{tabular}
\vspace{0.5cm}
\caption{Test Instances Based on the Belgian Network.} 
\label{Table:TestInstances}
\end{table}

Instances $B_1-B_4$ are based on the ``optimization from scratch''
benchmarks from \citep{Ref_DeWolfAndSmeers2000} and
\citep{Ref_BabonneauEtAl2012} ($\alpha=1,1.6,5,and 6$,
respectively). In these papers, the authors use the Belgian gas
network for a variation of the GTNEP problem which considers
integrated functions of the gas merchant and transportation
process. These benchmarks specify minimum and maximum production
levels (see Table~\ref{Table:LoadProfiles}). Since the GTNEP assumes
known gas nomination and production profiles, we computed load and
compression profiles based on optimal pressures provided in
\citep{Ref_BabonneauEtAl2012}.  Our instances also employ the same
cost function as in \citep{Ref_BabonneauEtAl2012} to compute the
associated costs for building new pipelines, i.e.,
\begin{equation*}
L_{ij}\left( 1.04081^{-6} D_{ij}^{2.5} + 11.2155\right)
\end{equation*}
where $D_{ij}$ and $L_{ij}$ are the diameter and length of pipeline
$(i,j)$ respectively. \citep{Ref_BabonneauEtAl2012} assumed continuous
diameter choices. However, we used a discrete diameter values
corresponding to the solution of \citep{Ref_DeWolfAndSmeers2000} and
Table 4 of \citep{Ref_BabonneauEtAl2012}.  For completeness, the
diameter choices are described in Table
\ref{Table:DiameterChoices}. Note that the exclusive-set constraint is
slightly different for these cases due to the existence of pre-defined
parallel pipes.  Within in each row of Table
\ref{Table:DiameterChoices}, the solution must contain one and only
diameter choice, and each set of parallel pipes must choose diameters
from the same column of Table \ref{Table:DiameterChoices}.

\begin{table}[t]
\centering
\begin{tabular}{cccccc}
\hline
Pipe & $D_1$ & $D_2$ & $D_3$ & $D_4$ & $D_5$\\ 
\hline
(1,2) A & 890.0 & 650.3 & 610.8 & 524.7 & 512.1  \\
(1,2) B & 890.0 & 650.3 & 610.8 & 524.7 & 512.1  \\
(2,3) A & 890.0 & 834.7 & 784.0 & 673.5 & 657.3 \\
(2,3) B & 890.0 & 834.7 & 784.0 & 673.5 & 657.3 \\
(3,4) & 890.0 & 998.9 & 938.3 & 806.0 & 786.7 \\
(5,6) & 590.1 & 604.3 & 567.6 & 487.6 & 475.9 \\
(6,7) & 590.1 & 0       & X       & X       & X     \\
(7,4) & 590.1 & 671.7 & 630.9 & 542.0 & 529.0 \\
(4,14) & 890.0 & 829.9 & 779.5 & 669.7 & 653.6 \\
(8,9) A & 890.0 & 902.8 & 848.0 & 728.4 & 711.0 \\
(8,9) B & 395.5 & 902.8 & 848.0 & 728.4 & 711.0 \\
(9,10) A & 890.0 & 902.8 & 848.0 & 728.4 & 710.9 \\
(9,10) B & 395.5 & 902.8 & 848.0 & 728.4 & 711.0 \\
(10.11) A & 890.0 & 787.6 & 739.8 & 635.5 & 620.1 \\
(10.11) B & 395.5 & 787.6 & 739.8 & 635.5 & 620.4 \\
(11,12) & 890.0 & 979.8 & 920.3 & 790.6 & 771.6 \\
(12,13) & 890.0 & 915.1 & 859.6 & 738.4 & 720.7 \\
(13,14) & 890.0 & 952.6 & 894.7 & 768.6 & 750.1 \\
(14,15) & 890.0 & 1201.0 & 1128.0 & 969.0 & 945.8 \\
(15,16) & 890.0 & 1038.4 & 975.3 & 837.9 & 817.7 \\
(11,17) & 395.5 & 469.0 & 440.5 & 378.4 & 369.3 \\
(17,18) & 315.5 & 469.0 & 440.5 & 378.4 & 369.3 \\
(18,19) & 315.5 & 469.0 & 440.5 & 378.4 & 369.3 \\
(19,20) & 315.5 & 448.9 & 421.7 & 362.2 & 353.5 \\
 \hline
\end{tabular}
\vspace{0.5cm}
\caption{Pipe diameter choices from Table 4 of \citep{Ref_BabonneauEtAl2012} } 
\label{Table:DiameterChoices}
\end{table}

\begin{table}[t]
\centering
\begin{tabular}{lp{.8cm}lp{.8cm}ccccc}
\hline\hline
              &   & & \multicolumn{3}{c}{(Loads)} & & \multicolumn{2}{c}{(Pressure)} \\
\small Node (Loc.) & \small Type$^{(*)}$ & & $\underline{L}$ & $\overline{L}$ & L  & &  $\underline{P}$ & $\overline{P}$ \\ \cline{1-2} \cline{4-6} \cline{8-9}\\
\small 1 (Zeebrugge)  & \small ${\cal I}$ & & 8.87 & 11.594 & 10.911288 &  & 0 & 77 \\
\small 2 (Dudzele)    & \small ${\cal I}$ & & 0 & 8.4 & 8.4 & & 0 & 77 \\
\small 3 (Brugge)     & \small ${\cal D}$ & & $-\infty$ & -3.918 & -3.918 & & 30 & 80 \\
\small 4 (Zomergem)   & \small   & & 0 & 0 & 0 & & 0 & 80 \\
\small 5 (Loenhout)   & \small ${\cal I}$ & & 0 & 4.8 & 2.814712 & & 0 & 77 \\
\small 6 (Antwerp)    & \small ${\cal D}$ & & -$\infty$ & -4.034 & -4.034 & & 30 & 80 \\
\small 7 (Ghent)      & \small ${\cal D}$ & & -$\infty$ & -5.256 & -5.256 & & 30 & 80 \\
\small 8 (Voeren)     & \small ${\cal I}$ & & 20.34 & 22.01 & 22.012 & & 50 & 66.2 \\
\small 9 (Berneau)    & \small   & &  0 & 0 & 0 & & 0 & 66.2$^\dagger$ \\
\small 10 (Li\`ege)      & \small ${\cal D}$ & & -$\infty$ & -6.365 & -6.365 & & 30 & 66.2 \\
\small 11 (Warnand)   & \small   & & 0 & 0 & 0 &  & 0 & 66.2 \\
\small 12 (Namur)     & \small ${\cal D}$ & & -$\infty$ & -2.12 & -2.12 & & 0 & 66.2 \\
\small 13 (Anderlues) & \small ${\cal I}$ & & 0 & 1.2 & 1.2 & & 0 & 66.2 \\
\small 14 (P\'eronnes)   & \small ${\cal I}$ & & 0 & 0.96 & 0.96 & & 0 & 66.2 \\
\small 15 (Mons)      & \small ${\cal D}$ & & -$\infty$ & -6.848 & -6.848 &  & 0 & 66.2 \\
\small 16 (Blaregnies)& \small ${\cal D}$ & & -$\infty$ & -15.616 & -15.616 & & 50 & 66.2 \\
\small 17 (Wanze)     & \small   & & 0 & 0 &0 &  & 0 & 66.2 \\
\small 18 (Sinsin)    & \small   & & 0 & 0 & 0 & & 0 & 63 \\
\small 19 (Arlon)     & \small ${\cal D}$ & & -$\infty$ & -0.222 & -0.222 & & 0 & 66.2 \\
\small 20 (P\'etange)    & \small ${\cal D}$ & & -$\infty$ & -1.919 & -1.919 & & 25 & 66.2 \\
 \hline
\end{tabular}
\vspace{0.5cm}
\caption{The Belgian Gas Network Data from ~\citep{Ref_DeWolfAndSmeers2000}. This data is used for the A Problems. $\dagger$- On Problems A1, A2, and A3,the pressure bounds are $[0,59.851968]$, $[0,59]$, and  $[0,59.85]$ respectively.} 
\label{Table:BelgianNetwork}
\end{table}

\begin{table}[t]
\centering
\begin{tabular}{clrrrr}
\hline\hline
          & & \multicolumn{4}{c}{Load ($L$) profiles (MMscf)}\\ \cline{3-6}
     Node & & \multicolumn{1}{c}{$\;\;\;\;B_1$} & \multicolumn{1}{c}{$\;\;\;\;B_2$} & \multicolumn{1}{c}{$\;\;\;\;B_3$} & \multicolumn{1}{c}{$\;\;\;\;B_4$} \\ \cline{1-1} \cline{3-6}
\small 1  & &   9.5883 &  9.8225 &   9.8218 &   9.7205 \\
\small 2  & &   8.1833 &  8.3447 &   8.1340 &   8.3628 \\
\small 3  & & -3.9180  & -3.9180 &  -3.9180 &  -3.9180 \\
\small 4  & &   0.0000 &  0.0000 &   0.0000 &   0.0000 \\
\small 5  & &   4.0315 &  4.0432 &   4.0383 &   4.0364 \\
\small 6  & &  -4.0315 & -4.0432 &  -4.0383 &  -4.0364 \\
\small 7  & &  -5.2413 & -5.2644 &  -5.2562 &  -5.2644 \\
\small 8  & &  22.012  & 22.0120 &  22.0120 &  22.0120 \\
\small 9  & &   0.0000 &  0.0000 &   0.0000 &   0.0000 \\
\small 10 & &  -6.4744 & -6.4951 &  -6.3970 &  -6.3816 \\
\small 11 & &   0.0000 &  0.0000 &   0.0000 &   0.0000 \\
\small 12 & &  -2.1929 & -2.1191 &  -2.1162 &  -2.0984 \\
\small 13 & &   1.2162 &  1.3225 &   1.0802 &   1.1591 \\
\small 14 & &   0.9840 &  0.6164 &   1.0776 &   1.0235 \\
\small 15 & &  -6.4056 & -6.5885 &  -6.8366 &  -6.8857 \\
\small 16 & & -15.6119 &-15.5904 & -15.4616 & -15.5899 \\
\small 17 & &  0.0000  &  0.0000 &   0.0000 &   0.0000 \\
\small 18 & &  0.0000  &  0.0000 &   0.0000 &   0.0000 \\
\small 19 & & -0.2059  & -0.2312 &  -0.2269 &  -0.2164 \\
\small 20 & & -1.9337  & -1.9112 &  -1.9131 &  -1.9236 \\ \hline
\end{tabular}
\vspace{0.5cm}
\caption{The Load Profiles Computed from Optimal Pressures Provided in~\citep{Ref_BabonneauEtAl2012}. All compression ratios were derived as 1.0.} 
\label{Table:LoadProfiles}
\end{table}

\begin{table}[t]
\centering
\small
\begin{tabular}{p{.65cm}lcccp{.6cm}p{.6cm}}
\hline\hline
Node& & Town & Lat. & Long. & $\underline{P}^{(*)}$ & $\overline{P}^{(*)}$ \\ \cline{1-1} \cline{3-7}
 & & \emph{Instance $A_1$}\\
21 & & Bois& 50.400676 & 5.855991 & 14 & 66 \\ 
22 & & Koninklijke & 50.806672 & 4.481877 & 14 & 66 \\
 & & \emph{Instance $A_2$}\\
21 & & Heist & 51.095651 & 4.744616 & 20 & 70 \\ 
22 & & Zoutleeuw & 50.858734 & 5.115404 & 20 & 70 \\
23 & & Beaufays & 50.552195 & 5.670182 & 20 & 70 \\
24 & & Gouvy & 50.231757 & 5.966813 & 20 & 70 \\
25 & & Ettelbruck & 49.861370 & 6.073930 & 20 & 70 \\
 & & \emph{Instance $A_3$}\\
21 & & Jabbeke & 51.204699 & 3.086440 & 14 & 66 \\
22 & & Torhout & 51.072867 & 3.118026 & 14 & 66 \\
23 & & Kortrijk & 50.790711 & 3.230636 & 14 & 66 \\
24 & & Bois-de-Barry & 50.580151 & 3.521773 & 14 & 66 \\ 
25 & & Lobbes & 50.353208 & 4.263261 & 20 & 70 \\
26 & & Senzeille & 50.124840 & 4.433550 & 20 & 70 \\ 
27 & & Gedinne & 49.980230 & 4.851030 & 20 & 70 \\
28 & & Chiny & 49.806832 & 5.274004 & 20 & 70 \\ 
29 & & Pigneule & 49.735878 & 5.471758 & 20 & 70 \\
  \cline{1-1} \cline{3-7}
\end{tabular}
\vspace{0.5cm}
\caption{Locations of Nodes of the Expansion Plans for the Belgian Gas Network. These nodes do not have injections.} 
\label{Table:NewNodes}
\end{table}

\begin{table}[t]
\centering
\small
\begin{tabular}{ccccp{.6cm}cc}
\hline\hline
Node & Node &  &  $w$ & $c$\\
\cline{1-2}  \cline{4-5} \\
\multicolumn{2}{c}{\emph{Instance $A_1$}}\\
9 & 21 & & 0.929 & 67.19\\
21 & 18 & & 0.808 & 77.26 \\
6 & 22 & & 0.785 & 79.50\\
22 & 14 & & 0.766 & 81.44\\
\multicolumn{2}{c}{\emph{Instance $A_2$}}\\
5 & 21 & & 1.052 & 59.29 \\ 
21 & $21^*$ & & Compressor & 1500.0\\ 
22 & 11 & & 0.967 &  64.52 \\ 
8 & 23 & & 1.933 & 32.28 \\ 
23 & 24 & & 0.876 &  71.18 \\ 
24 & $24^*$ & & Compressor & 1500.0 \\ 
25 & 19 & & 1.339 & 46.59 \\ 
$21^*$ & 22 & & 0.980 & 63.65 \\ 
$24^*$ &  25 & & 0.866 & 72.08 \\ 
\multicolumn{2}{c}{\emph{Instance $A_3$}}\\
1 & 21 && 2.257& 27.65\\ 
2 & 21 && 4.546 & 13.73\\ 
21 & $21^*$ && Compressor & 1500.0 \\ 
22 & 23 && 1.121 & 55.66\\ 
23 & $23^*$ && Compressor & 1500.0 \\ 
24 & 15 && 1.073 & 58.14\\ 
15 & 25 && 1.483 & 42.09 \\ 
25 & 26 && 1.289& 48.40\\ 
26 & $26^*$ && Compressor & 1500.0 \\ 
27 & 28 &&1.010 & 61.79\\ 
28 & 29 && 2.232&27.96 \\ 
29 & 19 && 1.423& 42.09\\ 
$21^*$ & 22 && 2.448& 25.50\\ 
$23^*$ & 24 &&1.165 & 53.56\\ 
$26^*$ & 27 && 1.071& 58.28\\ 
\hline
\end{tabular}
\vspace{0.5cm}
\caption{Locations of Pipes of the Expansion Plans for the Belgian Gas Network.  * denotes introduced dummy node for 0 length compressor arcs. } 
\label{Table:NewPipes}
\end{table}

\subsubsection{Larger Networks}

Table~\ref{Table:TestInstancesForScalability} describes the main data
points for the larger benchmarks. Instance $D$ is a real-life network
case whose data is restricted for confidentiality reasons and we are
not allowed to disclose its map or load profile. Instances $E, F$ and
$G$ are part of a German network whose data, including the network
configuration, maps, and load profiles, can be found
in~\citep{Ref_PfetschEtAl2012}.

\begin{table}[t]
\centering
\begin{tabular}{clcccccrcc}
\hline\hline
    & & \multicolumn{8}{c}{Network configuration} \\ [0.5ex] 
 \cline{3-10}
    & & & &  & \multicolumn{2}{c}{Base} & & \multicolumn{2}{c}{New} \\ \cline{6-7} \cline{9-10}
Ref. & & $|{\cal N}|$ & $|{\cal I}|$ & $|{\cal D}|$ & $|{\cal P}_e|$ & $|{\cal C}_e|$ & & $|{\cal P}_n|$ \\
\cline{1-1} \cline{3-7} \cline{9-10}
 $D$ & &  60 &  2 &  24 &  55 &  4 & &  55 \\
 $E$ & &  40 &  3 &  29 &  39 &  6 & &  39 \\
 $F$ & & 135 &  6 &  99 & 141 & 29 & & 141 \\
 $G$ & & 582 & 31 & 129 & 609 &  5 & & 278 \\
 \hline
\end{tabular}
\vspace{0.5cm}
\caption{Larger Instances of Gas Networks.} 
\label{Table:TestInstancesForScalability}
\end{table}

\subsection{The Algorithms and the Experimental Setting}
\label{subsection:algorithms}

This section reports computational results for three approaches:
\begin{enumerate}
\item The MINLP formulation of the GTNEP as shown in Model \ref{mod:MINLP};

\item The MISOCP relaxation of the GTNEP as shown in Model
  \ref{mod:MISOCP} followed by the conversion presented in Section
  \ref{section:conversion};

\item A MIP formulation based on a Piecewise Linear Approximation
  (PLA-MIP) of the quadratic functions; The PLA-MIP formulation
  follows the derivation
  in~\citep{Ref_CorreaPosadaAndSanchezMartin2014,Ref_DeWolfAndSmeers2000}
  and uses 60 segments.
\end{enumerate}

\noindent
All the experiments weere conducted on a computer with two Intel Xeon
CPU X5670 processors (2.93GHz) with 6 cores each. The computer has 64
GB DIMM 1333MHz RAM and runs the Ubuntu 14.04 LTS operating system.
The MINLP formulation is solved using SCIP
3.1.1~\citep{Ref_Achterberg2009} compiled with Ipopt 3.12.3 and Cplex
12.6. The PLA-MIP formulation is solved using CPLEX
12.6~\citep{Ref_Cplex2012}.  The MISCOP formulation is solved with
CPLEX 12.6 and the conversion is performed by IPOPT 3.12.3
\citep{Ref_WachterAndBiegler2006}.

\subsection{Results on the Belgian Network}
\label{subsection:belgium}

Table~\ref{Table:MathProperties} shows the sizes of the underlying
models in terms of the number of binary and continuous variables and
the number of linear and quadratic constraints for each instance.
Table~\ref{Table:ComputationalResults} presents the computational
results and reports the CPU time in seconds and the upgrade cost in
$\$\times10^3$ for each approach. The computational results show that
the MISOCP approach outperforms both the MINLP and the PLA-MIP and
that the solution to the MISOCP always converts to a feasible and
optimal solution. The PLA-MIP approach has both computational and
accuracy issues, as it significantly underestimates the optimal
objective value and is rather slow.

The results for problems B1--B4 are interesting as the expansion costs
are considerably lower than reported by \cite{Ref_BabonneauEtAl2012}
for the same operating conditions. In Table 4 of their paper,
\cite{Ref_BabonneauEtAl2012} report expansion costs of 15669, 14252,
11610, and 11274 for B1--B4. Their solutions are feasible and have the
same operating cost as our model. Of course, it is important to note
than their solutions were obtained through a model that minimizes
operating and expansion costs, which could make it harder to determine
the best design for particular operating conditions. Still, this
comparison highlights the strengths of the formulation proposed in this
paper.

\begin{table*}[t]
\centering
\begin{tabular}{clccrcclccrcclccrcc}
\hline\hline
    & & \multicolumn{5}{c}{MINLP}
    & & \multicolumn{5}{c}{PLA-MIP}
    & & \multicolumn{5}{c}{MISOCP}\\ [0.5ex]
    \cline{3-7} \cline{9-13} \cline{15-19}
Bench. & & BV & CV & & LC & QC & & BV & CV & & LC & QC & & BV & CV & & LC & QC \\ \cline{1-1} \cline{3-4} \cline{6-7} \cline{9-10} \cline{12-13} \cline{15-16} \cline{18-19}
 $A$   & &  54 & 49 & &254 & 96 &      & 1494 & 1837 & & 3931 & 0 &       &  54 & 73 & &  398 & 24 \\
 $A_1$ & &  70 & 59 & &320 & 112&      & 1750 & 2151 & & 4605 & 0 &       &  66 & 91 & &  488 & 28 \\
 $A_2$ & & 85 & 69 & &389 &124 &      & 1945 & 2391 & & 5120 & 0 &       & 78 & 107 & & 575 & 31 \\
 $A_3$ & & 103 & 80 & &463 &144 &      & 2263 & 2776 & & 5954 & 0 &       & 91 & 128 & & 679 & 36 \\
 $B_{1,2,3,4}$   & & 354 & 1154 & &464 &357 &      & 7314 &8737 & &19067 & 0 &       & 238 & 373 & & 1850 & 116 \\
 \hline
\end{tabular}
\vspace{0.5cm}
\caption{The Sizes of the Mathematical models for Belgian Network Instances. (BV: Binary variables, CV: Continuous variables, LC: Linear constraints, QC: Quadratic constraints). }
\label{Table:MathProperties}
\end{table*}

\begin{table*}[t]
\centering
\small
\begin{tabular}{clrrlrrlrr}
\hline\hline
     & & \multicolumn{2}{c}{MINLP}
     & & \multicolumn{2}{c}{PLA-MIP}
     & & \multicolumn{2}{c}{MISOCP}\\
     \cline{3-4} \cline{6-7} \cline{9-10} 
Bench. & & CPU & Obj & & CPU & Obj & & CPU & Obj \\
\cline{1-1} \cline{3-4} \cline{6-7} \cline{9-10} 
 $A$   & & 0.02 & 0.0 & & 0.6 & 0.0 & & 0.03 & 0.0 \\
 $A_1$ & & 0.06 & 144 & & 0.7 & 144 & & 0.05 & 144 \\
 $A_2$ & & 0.06 & 1687 & & 1.4 & 187 & & 0.1 & 1687 \\
 $A_3$ & & 0.06 & 1780 & & 1.9 & 280 & &  0.06 & 1780 \\
$B_1$ & & 1.89 & 11181 & & 1089 & 10353 & & 0.3 & 11181 \\
$B_2$ & & 3.17 & 11181 & & 1781 & 10361 & & 0.6 & 11181 \\
$B_3$   & & 3.53 & 11181 & &  1538 & 10352 & &  0.6 & 11181 \\
$B_4$   & & 3.82 & 11181 & &  1570 & 10352 & &  0.3 & 11181 \\
\cline{1-1} \cline{3-4} \cline{6-7} \cline{9-10}
\end{tabular}
\vspace{0.5cm}
\caption{Computational Results on the Belgian Network Instances: The Objective Value is in $\$$ and the CPU Time in 
Seconds. } 
\label{Table:ComputationalResults}
\end{table*}

\subsection{Scalability Results}
\label{subsection:scalability}

We now study whether the results on the Belgian networks continue to
hold on larger instances. To assess scalability and robustness, we
stress the networks by gradually increasing the production and
consumption levels from 5\% up to 300\% while considering solely the
addition of a parallel pipe for each existing pipeline in the base
configuration of the gas systems (i.e., $|{\cal C}_n| = 0$).  Table
\ref{Table:MathProperties} presents the sizes of the mathematical
models. In all of these results, we denote whether or not the MISOCP
and MINLP solutions are exact, lower bounds, or upper bounds on the
MINLP solutions. Lower bounds for the MINLP are also derived by
subtracting the optimality gap from any primal feasible solution.

Table \ref{Table:ComputationalResultsForScalabilityD} presents the
computational results on instance D which is based on proprietary
natural gas network in the United States.  Observe that the PLA-MIP
model systematically underestimates the objective function and returns
infeasible solutions. As we will see, this is systematic on all larger
benchmarks. The MISOP approach returns optimal solutions for all but
one case. Both the MINLP and MISOCP prove infeasibility of the most
stressed network.

Table \ref{Table:ComputationalResultsForScalabilityE} presents the
computational results on instance E which is based on gaslib-40
\citep{Ref_GasLib2014}.  The MISOP approach returns optimal solutions,
or proves infeasibilities in all cases. The MISOP model is one order
of magnitude faster than the MINLP model.

Table \ref{Table:ComputationalResultsForScalabilityF} presents the
computational results on instance F, which is based on gaslib-135
\citep{Ref_GasLib2014} and is particularly challenging. The MINLP
approach finds optimal solutions up to the 25\% case and spends
considerable time doing so. It finds an upper bound to the 50\% case
but does not return any information on the 75\% and 100\% cases.  In
contrast, the MISOCP approach finds optimal solutions to the 0\%, 5\%,
25\%, and 50\% cases, all below 10 seconds, It finds lower bounds on
the 75\% and 100\% cases reasonably fast. Both the MINLP and the
MISOCP prove infeasibility of the three most stressed instances.

Table \ref{Table:ComputationalResultsForScalabilityG} presents very
interesting results for instance G, which is based on gaslib-582
\citep{Ref_GasLib2014}. The MINLP approach cannot find feasible
solutions on any of the cases but the 300\% case which is shown
infeasible. Both the MINLP and PLA-MIP approaches have numerical
issues with these problems. The MISOP approach finds optimal solutions
up to the 50\% case and for the 150\% case and proves infeasibilities
for the 200\% and 300\% cases. For the 75\%--125\% cases, the MISCOP
times out but returns upper bounds to the optimal solution with
duality gaps ranging from 7.65\% to 51.3\%.

Overall, these results demonstrate the benefits of the MISOCP approach. 
The MISOCP approach almost always finds optimal solutions much faster
than the MINLP when both return optimal solutions. It also finds optimal
solutions or proves infeasibility in many case for the larger benchmarks,
while the MINLP approach does not return feasible solutions.

\begin{table*}[t]
\centering
\begin{tabular}{clccrcclccrcclccrcc}
\hline\hline
    & & \multicolumn{5}{c}{MINLP}
    & & \multicolumn{5}{c}{PLA-MIP}
    & & \multicolumn{5}{c}{MISOCP)}\\ [0.5ex]
    \cline{3-7} \cline{9-13} \cline{15-19}
 Ref. & & BV & CV & & LC & QC & & BV & CV & & LC & QC & & BV & CV & & LC & QC \\\cline{1-1} \cline{3-4} \cline{6-7} \cline{9-10} \cline{12-13} \cline{15-16} \cline{18-19}
 $D$ & & 283 & 174 & & 1093 & 440 &      & 6883 & 8330 & &  18018 & 0 &       & 228 &  339 & & 1753 & 110 \\
 $E$ & & 207 &  124 & & 792 & 312 &      & 4887 & 5920 & &  12796 & 0 &       & 168 &  241 & & 1260 &  78 \\
 $F$ & & 763 & 446 & & 2886& 1128 &      & 17683 & 21430 & & 46304 & 0 &       & 622 & 869 & & 4578 & 282  \\
 $G$ & & 2101 & 1469 & & 8058 & 2256 &      & 35941 & 44433 & & 92848 & 0 &       & 1823 & 2311 & & 11442 & 564 \\
 \hline
\end{tabular}
\vspace{0.5cm}
 \caption{Size of the Mathematical Models: BV: Binary variables, CV: Continuous variables, LC: Linear constraints, QC: Quadratic constraints.}
\label{Table:MathPropertiesForScalability}
\end{table*}

\begin{table*}[t]
\centering
\small
\begin{tabular}{rlrrlrrlrr}
\hline\hline
Stresss & & \multicolumn{2}{c}{MINLP} & & \multicolumn{2}{c}{PLA-MIP} & & \multicolumn{2}{c}{MISOCP} \\
\cline{3-4} \cline{6-7} \cline{9-10}  
level  & & CPU & Obj & & CPU & Obj & & CPU & Obj \\
\cline{1-1} \cline{3-4} \cline{6-7} \cline{9-10}  
0\%	  & & 0.1 & 0.00$^\bigstar$ & & 3.0 & 0.00 & &  0.1 & 0.00$^\bigstar$ \\
5\%	  & & 0.5 & 3.50$^\bigstar$ & & 1.8 & 0.00 & &  0.6 & 3.50$^\bigstar$ \\
10\%  & & 1.6 & 23.83$^\bigstar$ & & 12.2 & 23.22 & &  0.5 & 23.83$^\bigstar$ \\
25\%  & & 2.1 & 92.24$^\bigstar$ & & 14.0 & 83.99 & & 0.6 & 92.24$^\bigstar$  \\
50\%  & & 1.5 & 145.58$^\bigstar$ & & 14.8 & 136.2 & & 0.5 & 145.58$^\bigstar$ \\
75\%  & & 0.6 & 191.80$^\bigstar$ & & 11.0 & 184.0 & & 0.6 & 191.8$^\bigstar$ \\
100\% & & 3.0 & 287.00$^\bigstar$ & & 12.5 & 209.03 & & 0.7 & 281.99$\bigtriangleup$ \\
125\% & & 0.2 & $\dagger \; \; \; \;$ & & 1.6 & $\dagger \; \; \; \;$ & & 0.2 & $\dagger \; \; \; \;$ \\
\cline{1-1} \cline{3-4} \cline{6-7} \cline{9-10}  
\end{tabular}
\vspace{0.5cm}
\caption{Computational Results on Instance D. Obj: $\$\times 10^6$, CPU time: in seconds, Solution status: $^\bigstar$ = Proven optimal; $\bigtriangleup$ = Lower bound; $\bigtriangledown$ = Upper bound; $\dagger$ = Infeasible; $\ddagger$ = Unknown. }
\label{Table:ComputationalResultsForScalabilityD}
\end{table*}

\begin{table*}[t]
\centering
\small
\begin{tabular}{rlrrrlrrrlrrr}
\hline\hline
Stress  & & \multicolumn{3}{c}{MINLP} &    & \multicolumn{3}{c}{PLA-MIP} & & \multicolumn{3}{c}{MISOCP}\\ 
\cline{3-5} \cline{7-9} \cline{11-13}
level & & CPU & Obj & Gap & & CPU & Obj & Gap & & CPU & Obj & Gap \\ 
\cline{1-1} \cline{3-5} \cline{7-9} \cline{11-13}
  0\% & &   1.6 &   0.00$^\bigstar$ & 0.0 & & 10.2 & 0.00 & 0.0 & &  0.2 &  0.00$^\bigstar$ & 0.0 \\
  5\% & &   6.3 &  11.92$^\bigstar$ & 0.0 & & 23.5 & 0.00 & 0.0 & &  0.7 &  11.92$^\bigstar$  & 0.0 \\
 10\% & &  6.8 &  32.83$^\bigstar$ & 0.0 & & 20.6 & 0.00 & 0.0 & &  0.4 &  32.83$^\bigstar$  & 0.0 \\
 25\% & &  5.6 &  41.08$^\bigstar$ & 0.0 & & 30.9 & 32.8 & 0.0 & &  0.6 &  41.08$^\bigstar$  & 0.0 \\
 50\% & & 8.1 & 156.06$^\bigstar$ & 0.0 & & 11.5 & 32.8 & 0.0 & &  0.9 &  156.06$^\bigstar$  & 0.0 \\
 75\% & &  12.0 & 333.01$^\bigstar$ & 0.0 & & 21.8 & 121.1 & 0.0 & &  0.7 &  333.00$^\bigstar$  & 0.0 \\
100\% & &  12.1 & 551.64$^\bigstar$ & 0.0 & & 17.5 & 122.37 & 0.0 & &  0.8 &  551.64$^\bigstar$  & 0.0 \\
125\% & &  2.2 & $\dagger$ & -- & & 33.0 & 256.22 & 0.0 & &  0.4 & $\dagger$ & -- \\
150\% & &  0.8 & $\dagger$ & -- & & 27.6 & $\dagger$ & -- & & 0.3 & $\dagger$ & -- \\
 \cline{1-1} \cline{3-5} \cline{7-9} \cline{11-13}
\end{tabular}
\vspace{0.5cm}
\caption{Computational Results on Instance E. Obj: $\$\times 10^6$, CPU time: in seconds, Solution status: $^\bigstar$ = Proven optimal; $\bigtriangleup$ = Lower bound; $\bigtriangledown$ = Upper bound; $\dagger$ = Infeasible; $\ddagger$ = Unknown. }
\label{Table:ComputationalResultsForScalabilityE}
\end{table*}

\begin{table*}[t]
\centering
\small
\begin{tabular}{rlrrrlrrrlrrr}
\hline\hline
Stress  & & \multicolumn{3}{c}{MINLP} &    & \multicolumn{3}{c}{PLA-MIP} & & \multicolumn{3}{c}{MISOCP}\\ 
\cline{1-1} \cline{3-5} \cline{7-9} \cline{11-13}
level & & CPU & Obj & Gap & & CPU & Obj & Gap & & CPU & Obj & Gap \\ \cline{1-1} \cline{3-5} \cline{7-9} \cline{11-13}
  0\% & & 0.85 & 0.0$^\bigstar$ & 0.0 & & 136.3 &   0.0 & 0.0 & &   1.3 &   0.0$^\bigstar$ & 0.0 \\
  5\% & & 101.8 & 0.0$^\bigstar$ & 0.0 & & 120.0 &   0.0 & 0.0 & &   1.0 &   0.0$^\bigstar$ & 0.0 \\
 10\% & & 36707.3 & 15.04$^\bigstar$ & 0.0 & & 125.8 &   0.0 & 0.0 & &   2.4 &   0.0$\bigtriangleup$ & 0.0 \\
 25\% & & 457.9 & 60.4$^\bigstar$ & 0.0 & & 124.4 &   0.0 & 0.0 & &   4.4 &  60.4$^\bigstar$ & 0.0 \\
 50\% & & 86962.9 & 182.7$\bigtriangledown$ & 91.7 & & 166.7 &  60.4 & 0.0 & &   7.6 &  95.3$^\bigstar$ & 0.0 \\
 75\% & & 86933.9 & $\ddagger$ & -- & & 119.8 &  60.4 & 0.0 & &  40.5 & 451.5$\bigtriangleup$ & 0.0 \\
100\% & & 87334.2 & $\ddagger$ & -- & & 119.5 & 149.6 & 0.0 & & 104.6 &1234.2$\bigtriangleup$ & 0.0 \\
125\% & &   6.8 &  $\dagger$ & -- & & 125.7 & 149.6& 0.0 & &   1.8 & $\dagger$ & 0.0 \\
150\% & &    3.4 &  $\dagger$ & -- & & 206.7 & 486.0 & 0.0 & &   1.1 & $\dagger$ & 0.0 \\
200\% & &    0.4 &  $\dagger$ & -- & &  11.6 & $\dagger$ & -- & &   0.4 & $\dagger$ & 0.0 \\
 \cline{1-1} \cline{3-5} \cline{7-9} \cline{11-13}
\end{tabular}
\vspace{0.5cm}
\caption{Computational Results on Instance F. Obj: $\$\times 10^6$, CPU time: in seconds, Solution status: $^\bigstar$ = Proven optimal; $\bigtriangleup$ = Lower bound; $\bigtriangledown$ = Upper bound; $\dagger$ = Infeasible; $\ddagger$ = Unknown. }
\label{Table:ComputationalResultsForScalabilityF}
\end{table*}

\begin{table*}[t]
\centering
\small
\begin{tabular}{rlrrrlrrrlrrr}
\hline\hline
Stress  & & \multicolumn{3}{c}{MINLP} &    & \multicolumn{3}{c}{PLA-MIP} & & \multicolumn{3}{c}{MISOCP}\\ 
\cline{3-5} \cline{7-9} \cline{11-13}
level & & CPU & Obj & Gap & & CPU & Obj & Gap & & CPU & Obj & Gap \\ 
\cline{1-1} \cline{3-5} \cline{7-9} \cline{11-13}
  0\% & & 86400.0 & $\ddagger$ & -- & &  62012.9 &   6.87 & 0.0 & &   2.7 &   0.00$^\bigstar$  & 0.0 \\
  5\% & & 86400.0 & $\ddagger$ & -- & &  29655.1 &   2.78 & 0.0 & &   4.4 &   0.00$^\bigstar$  & 0.0 \\
 10\% & & 86400.0 & $\ddagger$ & -- & &  86400.0 &  4.65  & 40.22 & &  21.1 &   0.00$^\bigstar$  & 0.0  \\
 25\% & & 86400.0 & $\ddagger$ & -- & &  2153.2 &   8.65 & 0.0 & &  40.9 &   0.00$^\bigstar$ & 0.0  \\
 50\% & & 86400.0  & $\ddagger$ & -- & & 3670.2 &   $\dagger$ & -- & & 164.0 &   14.93$^\bigstar$ & 0.0  \\
 75\% & &  86400.0 & $\ddagger$ & -- & &  0.21 &  $\dagger$ & -- & & 86402.1 & 111.99$\bigtriangledown$ & 51.3 \\
100\% & & 86400.0 & $\ddagger$ & -- & & 5.31 & $\dagger$ & -- & & 86401.6 & 332.53$\bigtriangledown$ & 7.65  \\
125\% & & 86400.0 & $\ddagger$ & -- & & 5.31 & $\dagger$ & -- & & 86402.4 & 524.82$\bigtriangledown$ & 11.74  \\
150\% & & 86400.0  & $\ddagger$ & -- & &5.29 & $\dagger$ & -- & & 53321.3 & 590.84$^\bigstar$ & 0.0 \\
200\% & & 86400.0 & $\ddagger$ & -- & & 5.02 & $\dagger$ & -- & &  16.7 & $\dagger$ & --  \\
300\% & & 4.4 & $\dagger$ & -- & &   0.12 & $\dagger$ & -- & & 0.9 & $\dagger$ & --  \\
 \cline{1-1} \cline{3-5} \cline{7-9} \cline{11-13}
\end{tabular}
\vspace{0.5cm}
\caption{Computational Results on Instance G. Obj: $\$\times 10^6$, CPU time: in seconds, Solution status: $^\bigstar$ = Proven optimal; $\bigtriangleup$ = Lower bound; $\bigtriangledown$ = Upper bound; $\dagger$ = Infeasible; $\ddagger$ = Unknown. }
\label{Table:ComputationalResultsForScalabilityG}
\end{table*}

\subsection{The Importance of Integer Cuts}
\label{subsection:integercuts}

Table \ref{Table:ComputationalResultsWITHOUTCUTS} describes the
performance of the MISCOP on instances E, F, and G when the integer
cuts are not used. As can be seen, the integer cuts, which were used
both in the MINLP and MISOCP models, are critical to obtain an
efficient MISOCP implementation.

\begin{table*}[t]
\centering
\small
\begin{tabular}{rlrrrlrrrlrrr}
\hline\hline
Stress & & \multicolumn{3}{c}{Instance $E$} & & \multicolumn{3}{c}{Instance $F$} & & \multicolumn{3}{c}{Instance $G$}\\ \cline{3-5} \cline{7-9} \cline{11-13}
level & & CPU & Obj & Gap & & CPU & Obj & Gap & & CPU & Obj & Gap \\ 
\cline{1-1} \cline{3-5} \cline{7-9} \cline{11-13}
  0\% & &  0.9 &   0.00$^\bigstar$ & 0.0 & & 3310.9 &   0.00$^\bigstar$        &  0.0 & &  242.2 &   0.00$^\bigstar$ & 0.0 \\
  5\% & &  1.8 &  11.92$^\bigstar$ & 0.0 & &   83.5 &   0.00$^\bigstar$        &  0.0 & &   14.2 &   0.00$^\bigstar$ & 0.0 \\
 10\% & &  2.7 &  32.83$^\bigstar$ & 0.0 & &  120.7 &   0.00$\bigtriangleup$   &  0.0 & &  301.5 &  0.00$^\bigstar$ & 0.0 \\
 25\% & &  3.2 &  41.08$^\bigstar$ & 0.0 & & 86419.5 &  60.44$\bigtriangledown$ & 75.1 & & 86400.3 & $\ddagger$ & -- \\
 50\% & &  8.5 & 156.06$^\bigstar$ & 0.0 & & 17693.1 &  95.32$\bigstar$ & 0.0 & & 8271.05 &  14.93$^\bigstar$  & 0.0 \\
 75\% & &  6.7 & 333.01$^\bigstar$ & 0.0 & & 86409.9 & 451.59         & 59.2 & & 86404.4 &  111.99$\bigtriangledown$ & 79.4 \\
100\% & &  3.8 & 551.64$^\bigstar$ & 0.0 & & 86404.5 &1234.23         & 44.2 & & 87193.1 & 332.32$\bigtriangledown$ & 87.9 \\
125\% & &  1.8 & $\dagger$ & -- & & 90.9 & $\dagger$    &   -- & & 86401.8 & 524.82$\bigtriangledown$ & 16.1 \\
150\% & &  1.0 & $\dagger$ & -- & &  7.5 & $\dagger$    &   -- & & 86408.9 & 245.80$\bigtriangleup$ & 58.5 \\
200\% & &  0.7 & $\dagger$ & -- & &  2.0 & $\dagger$    &   -- & & 13318.6 & $\dagger$  & -- \\
300\% & &  0.0 & $\dagger$ & -- & &  0.0 & $\dagger$    &   -- & &    3.0 & $\dagger$  & --  \\
\cline{1-1} \cline{3-5} \cline{7-9} \cline{11-13}
\end{tabular}
\caption{Computational Results on Instances E, F, and G without the Integer Cuts. Obj: $\$\times 10^6$, CPU time: in seconds, Solution status: $^\bigstar$ = Proven optimal; $\bigtriangleup$ = Lower bound; $\bigtriangledown$ = Upper bound; $\dagger$ = Infeasible; $\ddagger$ = Unknown }
\label{Table:ComputationalResultsWITHOUTCUTS}
\end{table*}

\section{Concluding Remarks} 
\label{Sec:Conclusions}

This paper considered the expansion of natural gas networks, a
critical process involving substantial capital expenditures with
complex decision-support requirements. It proposed a convex
mixed-integer second-order cone relaxation for the gas expansion
planning problem under steady-state conditions in order to address the
fact that state-of-the-art global optimisation solvers are unable to
scale up to real-world size instances. The resulting MISOCP model
offers tight lower bounds with high computational efficiency. In
addition, the optimal solution of the relaxation can often be used to
derive high-quality solutions to the original problem, leading to
provably tight optimality gaps and, in some cases, global optimal
solutions. The convex relaxation is based on a few key ideas,
including the introduction of flux direction variables, exact
McCormick relaxations, on/off constraints, and integer cuts. Numerical
experiments are conducted on the traditional Belgian gas network, as
well as other real larger networks. The computational results
demonstrate that the MISOCP model is faster than the originating MINLP
model by one or two orders of magnitude on the Belgian network
instances. They also show that the MISOCP model scales well to large
and stressed instances.

\end{document}